\documentclass[useAMS, usenatbib, usegraphicx]{mn2e}
\usepackage{subfigure}
\usepackage{txfonts}
\usepackage[normalem]{ulem}
\usepackage{color}
\usepackage[usenames,dvipsnames]{xcolor}
\def\deg{\ifmmode^\circ\else$^\circ$\fi}
\def\arcs{\ifmmode {^{\scriptstyle\prime\prime}}
          \else $^{\scriptstyle\prime\prime}$\fi}
\def\arcm{\ifmmode {^{\scriptstyle\prime}}
          \else $^{\scriptstyle\prime}$\fi}
\def\ra[#1 #2 #3]{#1\sup{h}\,#2\sup{m}\,#3\sup{s}}
\def\dec[#1 #2 #3]{#1\deg\,#2\arcm\,#3\arcs}
\def\uJy{$\mu$Jy\,}
\def\mJy{mJy}
\def\vv#1{\ifmmode {\mathbf{#1}}\else ${\mathbf{#1}}$\fi}
\def\mcadam{\textsc{McAdam}}

\title[Redshift-one clusters]{AMI SZ observations and Bayesian analysis of a sample of six redshift-one clusters of galaxies\thanks{We request that any reference to this paper cites `AMI Consortium: Schammel et al.'} }
\author[AMI Consortium: Schammel et~al.]{AMI Consortium: Michel P. Schammel$^{1}$\thanks{E-mail:
m.schammel@mrao.cam.ac.uk}, Farhan~Feroz$^{1}$, Keith~J.~B.~Grainge$^{1,3}$, \newauthor
	 Michael~P.~Hobson$^{1}$,
   Natasha~Hurley-Walker$^{4}$, Anthony~N.~Lasenby$^{1,3}$, \newauthor
   Malak~Olamaie$^{1}$, Yvette~C.~Perrott$^{1}$, Guy~G.~Pooley$^{1}$,  \newauthor
   Carmen~Rodr\'{i}guez-Gonz\'{a}lvez$^{5}$, Clare~Rumsey$^{1}$, Richard~D.~E.~Saunders$^{1,3}$, \newauthor
   Paul~F.~Scott$^{1}$, Timothy~W.~Shimwell$^{2}$, David~J.~Titterington$^{1}$,\newauthor Elizabeth~M.~Waldram$^{1}$\\
$^1$ Astrophysics Group, Cavendish Laboratory,
     19 J.~J.~Thomson Avenue, Cambridge CB3 0HE \\
$^2$ CSIRO Astronomy \& Space Science, Australia Telescope National Facility, PO Box 76, Epping, NSW 1710, Australia \\
$^3$ Kavli Institute for Cosmology Cambridge,
     Madingley Road, Cambridge CB3 0HA\\
$^4$ International Centre for Radio Astronomy Research, Curtin Institute of Radio Astronomy, 1 Turner Avenue,\\ Technology Park, Bentley, WA 6845, Australia \\
$^5$ Spitzer Science Center, MS 220-6, California Institute of Technology, Pasadena, CA 91125 USA
\\
}

\begin{document}

\date{Accepted ---; received ---; in original form \today}

\pagerange{\pageref{firstpage}--\pageref{lastpage}} \pubyear{2012}

\maketitle

\label{firstpage}

\begin{abstract}
We present 16-GHz Sunyaev-Zel'dovich observations using the Arcminute Microkelvin Imager (AMI) and subsequent Bayesian analysis of six galaxy clusters at redshift $z~\approx~1$ chosen from an X-ray and Infrared selected sample from \citet{Culverhouse2010}. In the subsequent analysis we use two cluster models, an isothermal $\beta$-model and a Dark Matter GNFW (DM-GNFW) model in order to derive a formal detection probability and the cluster parameters.
We detect two clusters (CLJ1415+3612 \& XMJ0830+5241) and measure their total masses out to a radius of 200 $\times$ the critical density at the respective cluster's redshift. For CLJ1415+3612, we find  $M_{\mathrm{T},200} = 7.3^{+1.8}_{-1.8} \times 10^{14} M_\odot$ ($\beta$ model) and $M_{\mathrm{T},200} = 10.4_{-2.4}^{2.5} \times 10^{14} M_\odot$ (DM-GNFW model) and for XMJ0830+5241, we find $M_{\mathrm{T},200} = 3.6^{+1.1}_{-1.1} \times 10^{14} M_\odot$, ($\beta$-model) and $M_{\mathrm{T},200} = 4.7_{-1.4}^{+1.4} \times 10^{14} M_{\odot}$ (DM-GNFW model), which agree with each other for each cluster. We also present maps before and after source subtraction of the entire sample and provide 1D and 2D posterior marginalised probability distributions for each fitted cluster profile parameter of the detected clusters. 
Using simulations which take into account the measured source environment from the AMI Large Array (LA), source confusion noise, CMB primordials, instrument noise, we estimate from small-radius ($r_{2500}$) X-ray data from \citet{Culverhouse2010} the detectability of each cluster in the sample and compare it with the result from the Small Array (SA) data. Furthermore, we discuss the validity of the assumptions of isothermality and constant gas mass fraction. We comment on the bias that these small-radius estimates introduce to large-radius SZ predictions. In addition, we follow-up the two detections with deep, single-pointed LA observations. We find a $3$-$\sigma$ tentative decrement toward CLJ1415+3612 at high-resolution and a $5$-$\sigma$ high-resolution decrement towards XMJ0830+5241.
\end{abstract}

\begin{keywords}
 cosmology: observations -- cosmic microwave background --
 galaxies: clusters -- Sunyaev--Zel'dovich X-ray -- galaxies: clusters:
  individual (CLJ1415+3612, ISCS1438+34, RXJ0910+5422, SPJ1638+4039, XMJ0830+5241, XMJ0849+4452) 
\end{keywords}

\section{Introduction}

The Sunyaev\textendash{}Zel\textquoteright{}dovich (SZ) effect (Sunyaev
\& Zeldovich 1970, 1972) is the inverse-Compton scattering of the
CMB radiation by hot, ionised gas in the gravitational potential well
of a cluster of galaxies; for reviews on the SZ effect (see e.g. \citet{Birkinshaw1999} and \citet{Carlstrom2002}). The effect is useful in a number of ways for the study
of galaxy clusters; here we are concerned with two in particular.
First, because the SZ effect arises from a scattering process, a cluster
at one redshift will produce the same observed SZ surface brightness
as an identical cluster at any other redshift. On the other hand, the integrated SZ flux does depend on the angular diameter distance and therefore the redshift $z$, but the dependence is small at $z \gtrapprox 0.5 $. Hence, the usual sensitivity issue of high-redshift observing is avoided, which is particularly useful for this study. Second, since the
SZ surface brightness is proportional to the line-of-sight integral
of pressure through the cluster, the SZ signal is less sensitive to
concentration than the X-ray Bremsstrahlung signal; one corollary
of this is that the ratio SZ-sensitivity / X-ray-sensitivity increases
with distance from the cluster centre so that with SZ one can probe
out to, say, the virial radius, provided the telescope is sensitive
to sufficiently large angular scales. SZ decrements are faint, however,
and can be contaminated or obliterated by sources of radio emission.
Many SZ studies are currently and routinely carried out at different redshift ranges by the Atacama Cosmology Telescope (ACT) (see e.g. \citealt{Seghal2012}), the \textit{Planck} satellite (see e.g. \citealt{PlanckAMI2012}), the South Pole Telescope (SPT) (see e.g. \citealt{Stadler2012}) and the Sunyaev-Zel'dovich Array (SZA, see e.g. \citealt{Muchovej2007}).
The Arcminute Microkelvin Imager (AMI) has demonstrated its capability of studying X-ray selected clusters of galaxies at moderate $z$, (see e.g. \citet{Zwart2011}, \citet{Rodriguez-Gonzalvez2011}), however always at relatively low redshift. Here we investigate the feasibility of observing and detecting clusters at higher ($z\approx1$) redshift as well as constraining their masses.
\subsection{Sample selection}\label{sec:sample_selection}

\citet{Culverhouse2010} selected a sample of 11 clusters at ($z \approx 1$) from infrared and X-ray studies, and observed these with the SZA to search for SZ signal. They found SZ decrements towards 3 of the clusters; using X-ray based values of $r_{2500}$, they also estimated $Y_{2500}$ and $M_{gas,2500}$ (from the $Y$-$M_{gas}$ scaling relation of \citealt{Bonamente2008}). Here, we present AMI observations of the subsample of 6 (of the original 11) clusters which are at $\delta \ge 20\degr$, along with Bayesian inference where possible. Our initial aims were to:
\begin{enumerate}
\item follow up the SZA observations with an instrument with different systematics;
\item consider the effects of difference in analysis procedures used by \citet{Culverhouse2010} (from X-ray measurements near the cluster centre, and assuming hydrostatic equilibrium; these assumptions are often made in the literature) and by ourselves (temperature from SZ observation plus M-T relation from virial theorem, as well as Bayesian and probability search).
\end{enumerate} 
 The names, coordinates and redshifts of the sample are shown in Table \ref{tab:summary}.

In section \ref{sec:AMI} we briefly review AMI and then describe the observational programme in section \ref{sec:observations}.  
The data reduction and Bayesian analysis are explained in section \ref{sec:data_red}. We have conducted realistic AMI simulations using X-ray measured parameters to model the galaxy cluster and discuss their validity in section \ref{sec:simulations}, present the SA analysis in section \ref{sec:sa_results} and the LA follow-up of two clusters in section \ref{sec:la_followup}. Throughout, we assume $H_0 = 70$ km\,s$^{-1}$\,Mpc$^{-1}$ and a concordance $\Lambda$CDM cosmology
with $\Omega_{\rm{m},0}=0.3$, $\Omega_{\Lambda,0}=0.7$, $\Omega_k=0$,
$\Omega_{b,0}=0.041$, $w_0=-1$, $w_a=0$ and $\sigma_8=0.8$.

\section{The Arcminute Microkelvin Imager}\label{sec:AMI}

The Arcminute Microkelvin Imager (AMI) is a dual interferometric array, near Cambridge. It consists of an 8-element Large Array (LA) with dishes of 12.8m and the
Small Array (SA), a 10-dish array with an antenna size of 3.7m each. Both arrays observe at a central frequency of 15.75 GHz and their configuration is optimised for the specific scientific area they were conceived for. The main characteristics of AMI are summarised in Table \ref{tab:AMI_specs} and are described in greater detail in \citet{Zwart2008}. 

\begin{table}
\caption{AMI technical summary.} \label{tab:AMI_specs} 
\begin{tabular}{ccc}
 & { SA} & {LA}\tabularnewline \hline \hline  
Antenna diameter & {$3.7$\, m} & $12.8$\, m\tabularnewline 
{Number of antennas} & {10} & { 8}\tabularnewline 
{Baseline lengths (current)} & {$5-20$\,m} & {$18-110$\,m}\tabularnewline 
{Primary beam at $15.7$\,GHz} & {$20^{'}.1$} & { $5^{'}.5$}\tabularnewline 
{Synthesized beam} & {$\approx3^{'}$} & { $\approx30^{''}$}\tabularnewline 
{Flux sensitivity} & {$3\textrm{0 mJy s}^{-1/2}$} & { $3\textrm{ mJy s}^{-1/2}$}\tabularnewline 
{Observing frequency} & {$13.9-18.2$GHz} & { $13.9-18.2$GHz}\tabularnewline 
{Bandwidth} & {$4.3$\,GHz} & { $4.3$\,GHz}\tabularnewline  
{Number of channels} & {6} & {6}\tabularnewline 
{ Channel bandwidth} & { $0.72$\,GHz} & { $0.72$\,GHz}\tabularnewline 
\end{tabular} 
\end{table}
\begin{table*}
\caption{Cluster sample observed by AMI: redshifts, coordinates and the reference of the initial detection/redshift measurement}\label{tab:summary}
\begin{tabular}{ccccc}

Cluster name & Redshift & R.A. & DEC. & Detection \tabularnewline
\hline
\hline 
CLJ1415+3612 & 1.03 & 14 15 11 & 36 12 04 &  X-ray, \citet{Maughan2006}, \citet{Perlman2002}\\[0.3mm] 

ISCS1438+34 &  1.41 & 14 38 09  & 34 14 19 & IR, \citet{Stanford2005} \\[0.3mm] 

RDJ0910+5422 & 1.11 & 09 10 45 & 54 22 09 &  X-ray, \citet{Stanford2002} \\[0.3mm] 
 
SPJ1638+4039 & 1.20 & 16 38 52 & 40 38 43  & IR,  \citet{Muzzin2009} \\[0.3mm] 
 
XMJ0830+5241 & 0.99 & 8 30 26  & 52 41 33 &  X-ray, \citet{Lamer2008} \\[0.3mm] 

XMJ0849+4452 & 1.26 & 08 48 59  & 44 51 50 &  X-ray, \citet{Rosati1999} \\[0.3mm] 

\end{tabular}
\end{table*}

\begin{table}
\caption{Noise levels of each target on the central regions of the continuum maps and integration times on both SA and the corresponding LA raster observations.}\label{tab:observations}
\begin{tabular}{c|cc|cc}
Cluster & SA noise & LA noise & t$_\mathrm{obs, SA}$  & t$_\mathrm{obs, LA}$ \\
Name & (\uJy)& (\uJy) & (hours) & (hours)  \\
\hline
\hline
CLJ1415+3612 & 95 & 65 & 36 & 17  \\
XMJ0849+4452 & 70 & 80 & 11 & 22 \\
ISCS1438+34 & 90 & 75 & 71& 42 \\
RDJ0910+5422 & 100 & 115 & 34 & 12 \\
SPJ1638+4039 &  145 & 85 & 55 & 18 \\
XMJ0830+5241 & 70 & 45 & 44 & 31 \\

\end{tabular}
\end{table}

\section{AMI Observations}\label{sec:observations}

All the clusters were observed using both arrays; a mosaicing strategy was used to cover the same area with the LA as the single pointed observations measured by the SA. Integration times and map noise levels for each target on both arrays are shown in Table \ref{tab:observations}. Some of the clusters had substantially more integration time (e.g. ISCS1438+34) compared to others which reflects the amount of flagging necessary to remove interference and to assure a highly-filled circularly symmetric uv-coverage to sample all observable spatial scales and obtain a circular synthesised beam. All observations were taken between May 2008 and March 2011 on the SA, while integration on the individual targets were all carried out within a few months to reduce the effect of source variability. During each run on the SA we observed a close-by secondary calibrator for 400 seconds each 6 minutes to maintain phase stability on the SA. The LA mosaicing runs were carried out within a few days of each SA observation and each run had a secondary calibrator interleaved every 10 minutes for 2 minutes. Although the SA is our primary SZ array with baseline ranges geared to be more sensitive to SZ flux and larger scales, we followed-up two targets (CLJ1415+3612 and XMJ0830+5241) with deep LA single pointings, as high redshift clusters are expected to have a smaller angular extent due to the angular diameter distance--redshift relation. SZ observations at high redshift could therefore benefit from the increased resolution of the LA. We assess this by choosing the two SA detections (see section \ref{sec:sa_results}) and get further detection confirmation by carrying out these deep follow-up observations on each target within a month in 2012. LA integration time for the single pointings on CLJ1415+3612 was 8 hours in total and reached a sensitivity of 25 \uJy (50 \uJy on the shortest, $<1.5 \mathrm{k}\lambda$ baselines). For XMJ0830+5241 a sensitivity of 20 \uJy (45 \uJy on the shortest, $<1.5 \mathrm{k}\lambda$ baselines) was reached after 13 hours. The pointed LA observations visit the phase calibrator every 10 minutes for 100 seconds.

\section{Data Reduction and Analysis} \label{sec:data_red}
\subsection{Data reduction and mapping}

Individually, the data gathered from each run on both arrays are flagged for shadowing effects, slow fringe rates, pointing -- and path compensator delay errors with our in-house reduction package \textsc{reduce}.  Absolute flux calibration is performed using daily observed calibrators, 3C48 and 3C286. The data are then Fourier transformed and fringe-rotated to the pointing centre. Further flaggings reject interference and discrepant baselines. Phase calibration is performed using the interleaved calibrators. Last, the amplitudes of the visibilities are weighted to take into account system temperature variations, due to weather and airmass, before being outputted for every frequency channel as \textsc{UVFITS}. 

Our in-house software \textsc{fuse} is used to concatenate the \textsc{UVFITS} of each observation and then the data are mapped using the imaging package \textsc{aips}\footnote{http://www.aips.nrao.edu} for both the single frequency channels and the full spectrum. An initial deconvolution to estimate the real map noise is done using the \textsc{clean}-algorithm with a flux limit of 3 times the noise calculated from the weights for each channel and pointing centre. The noise level of those initial maps are calculated and serve as flux limit (3 $\times$ the calculated value)  for a second deconvolution to produce the final maps. We use a box encompassing the entire primary beam. For more details on the mapping technique, (see e.g. \citet{Franzen2011} and \citet{Shimwell2012}).

\subsection{Source finding}\label{sec:sourcefind}
The LA maps are used to perform source finding and spectral index fitting using our in-house software \textsc{sourcefind}. We give a brief summary of the technique here, for a more detailed description see \citet{Franzen2011}. All pixels on the map with a flux density greater than $0.6 \times \gamma \times \sigma_n$, where $\sigma_n$ is the noise map value for that pixel and $\gamma = 4$ is the desired detection threshold, are identified as peaks. The flux densities and positions of the peaks are determined using a tabulated Gaussian sinc degridding function to interpolate between the pixels and the peaks above a threshold of $\gamma \times \sigma_n $ are identified as sources. In addition the \textsc{aips} routine \textsc{jmfit} is used to fit a two-dimensional Gaussian to each source to give the angular size and the integrated flux density for the source. These fitted values are compared to the point-source response function of the telescope to determine whether the source is extended on the LA map. 

Assuming a power-law relationship between flux density and frequency ($S \propto \nu^{-\alpha}$) and given the measured flux of each individual channel map, a spectral index was calculated using an MCMC method called \textsc{metromod} \citep{Hobson2004} -- the prior on the spectral index has a Gaussian distribution with a mean of 0.5 and $\sigma$ of 2.0, truncated at $\pm 5.0$. The positions, fluxes and spectral indices are retained for use in our Bayesian analysis.

\subsection{Preparing SA data for Bayesian analysis}
The concatenated SA data usually contain more than 5~000 visibilities per channel which is not tractable for our Bayesian analysis package. The data therefore undergo a binning step using a bin width set to a fifth of the width of the aperture illumination function which leaves about 1000 visibilities per channel, evenly spaced in \textit{uv}-space, to be used for further analysis. The bin width was chosen in order to assure that enough samples populate each bin and no information will be lost during the process. 

\subsection{\mcadam~-- Bayesian Analysis} \label{sec:mcadam}
Our Bayesian Analysis package, \mcadam~\citep{Marshall2003} uses a fast sampler, \textsc{multinest} \citep{Feroz2009} and performs a joint fit of the \textit{uv}-data for the presence of a cluster imprint via a physical or analytical model and contaminations from undesired signals originating from CMB primordials, radio sources, instrumental noise and source confusion. This is done in a fully Bayesian way; the sampling takes into account all the prior knowledge of the cluster model and the source environment investigated by the LA in order to do its simultaneous fit.

Despite this efficient use of initial knowledge which considerably narrows down the size of the parameter space to be explored, the dimensionality of the problem (e.g. 40 dimensions in the case of ISCS1438+34), sometimes requires further simplification of the problem by fixing the flux and spectral index of sources whose integrated fluxes are less than $4 \sigma_{\mathrm{SA}}$, where $\sigma_{\mathrm{SA}}$ is the continuum noise level on the SA. 
For this analysis, we use two different models to describe and fit for the cluster and its imprint on the sky, as follows.

\subsubsection{$\beta$-model}\label{sec:beta-model}

First, we use an isothermal $\beta$-model \citep{Cavaliere1976, Cavaliere1978} which has proven itself before; see e.g. \citet{Hurley-Walker2012} for a comparison of mass derivations using an SZ $\beta$-model and weak-lensing data but also \citet{Rodriguez-Gonzalvez2011} and \citet{AMIOlamaie}. The model describes a cluster gas electron density $n_e$ which decreases with radius $r$ according to

\begin{equation}
\rho_g(r) = \frac{\rho_g(0)}{\left(1+\frac{r^2}{r^2_c}\right)^{3\beta/2}} \mathrm{,}
\end{equation}
where $\rho_g(0)$ is the gas mass per electron ($\rho_g(r) = n_e(r) \times 1.14 m_p$, $m_p$ is the proton mass) and $r_c$ is the core radius. We measure the total mass $M_{T,200}$ at radius $r_{200}$, the radius at which the mean enclosed density is 200 times the critical density $\rho_\mathrm{crit}$ at the cluster's redshift $z$ and assuming spherical symmetry:
\begin{equation}
M_{\mathrm{T},200} = \frac{4\pi}{3}r^3_{200} (200 \rho_\mathrm{crit}) \mathrm{.}
\end{equation}

Following \citet{Voit2005} and assuming the cluster is virialised at this radius, we can relate a collapsing top-hat density perturbation model to a singular truncated isothermal sphere. This also takes into 
account the finite boundary pressure and assumes all kinetic energy is internal 
energy of the hot plasma. Hence, we calculate the temperature at $r_{200}$ via a mass-temperature scaling relation
\begin{equation}\label{eq:MTrel}
 k_{\rm B}T_{\mathrm{g, 200}} =\frac{\mu}{2}\left(\frac{200}{2}\right)^{1/3} \times \left[GM_{\rm {T,200}} H(z)\right]^{2/3}.
\end{equation}

In summary, the data set can be fitted to a full cluster model with parameters $\Theta_c =  (x_c, y_c, \beta, r_c, M_\mathrm{T, 200}, f_\mathrm{gas,200})$ and radio source parameters $\Psi =(x_s, y_s, S_0, \alpha)$. A summary of the priors on each parameter is given in Table \ref{tab:priors}. [Note that the methodology works even if the redshift of the cluster is unknown; in such a case $z$ is simply appended to the list of parameters $\Theta_c$.]

\begin{table*}
\caption{Priors for the cluster and source parameters used for the Bayesian analysis of our data. We list the parameterisation of the $\beta$-model first, then the source parametersation and prior distributions for each individual object which are used for both cluster models. The last section lists the alternative parameters and priors when using the DM-GNFW cluster model.}\label{tab:priors}
\begin{tabular}{lcc}
Parameter & Prior used \\  \hline \hline
Cluster Position ($x_c$, $y_c$) & Gaussian at x$_\mathrm{cluster}$, $\sigma = 60 \arcsec$ \\
Core radius ($r_{c}$ /kpc)) & Uniform between 10 and 1000 \\ 
Beta   ($\beta$)  & Uniform between 0.3 and 2.5 \\ 
Mass ($M_{T,200}$ /$M_{\odot}$) & Uniform in log space between 3 $\times$ $10^{13}$ and 5 $\times$ $10^{15}$ \\ 
Gas fraction ($f_\mathrm{g,200}$) & Gaussian prior centred on 0.123 \citep{Larson2011, Zhang2010} with  $\sigma = 0.03$\\
\hline

Source position ($x_{s}, y_{s}$) & A delta-function prior using the LA positions \\ 
Source flux density ($S_{0}/\rm{Jy}$) & A Gaussian centred on the LA continuum value with a $\sigma$ of 40 per cent \\ 
Source spectral index ($\alpha$) & A Gaussian centred on the fitted spectrum and the LA error as $\sigma$  \\ 
\hline
Cluster Position ($x_c$, $y_c$) & Gaussian at x$_\mathrm{cluster}$, $\sigma = 60 \arcsec$ \\
Mass ($M_{T,200}$ /$M_{\odot}$) & Uniform in log space between 3 $\times$ $10^{13}$ and 5 $\times$ $10^{15}$ \\ 
Gas fraction ($f_\mathrm{g,200}$) & Gaussian prior centred at 0.1 with $\sigma = 0.02$\\
\end{tabular}
\end{table*}

\subsubsection{DM-GNFW model}
This parameterisation uses an analytical model described in \citet{Olamaie2012a} which models the dark matter halo of a cluster of galaxies with an NFW-profile \citep{Navarro1997} and a GNFW \citep{Nagai2007} pressure profile to describe the cluster gas. This model relies on 2 sample parameters, which are all given at $r_{200}$: M$_\mathrm{T, 200}$ and f$_\mathrm{gas}$.
We calculate the halo concentration parameter $c_{200}$ as a function of cluster mass and redshift calculated using a relaxed cluster relationship \citep{Olamaie2012b, Neto2007}:

\begin{equation}
c_{200} = \frac{5.26}{1+z} \times \left(\frac{M_{\mathrm{T,200}}}{10^{14} h^{-1} M_\odot}\right)^{-0.1}\mathrm{.}
\end{equation}
The lower section of Table \ref{tab:priors} shows the priors used for the DM-GNFW model analysis; note that the cluster's positional and redshift priors are identical to those used in the $\beta$-model. 
\subsubsection{Source parameters}
In both cases, the same source parameterisation is used; we fit their positions with a delta prior as the LA is able to measure positions to greater precision than the SA. All of the sources brighter than four times the noise level on the SA maps are fitted.

\subsection{Detection of a cluster}
Using Bayesian analysis carries the great advantage that we can compare two hypotheses, namely the existence of a cluster ($H_1$) in a particular search area and the hypothesis of there not being a cluster in the same area ($H_0$), using the ratio R:
\begin{equation}\label{eq:ev_ratio}
R = \frac{\mathrm{Pr}(H_1|D)}{\mathrm{Pr}(H_0|D)} = \frac{\mathrm{Pr}(D|H_1)}{\mathrm{Pr}(D|H_0)}\frac{\mathrm{Pr}(H_1)}{\mathrm{Pr}(H_0)} \mathrm{.}
\end{equation}
$R$ depends only on the ratio $\frac{\mathrm{Pr}(H_1)}{\mathrm{Pr}(H_0)}$ of the prior probabilities and the evidence ratio $\frac{\mathrm{Pr}(D|H_1)}{\mathrm{Pr}(D|H_0)}$, which is an output from \textsc{multinest}. The ratio of prior probabilities can normally be set to unity but occasionally requires further consideration. Hence a formal detection criterion can be derived by analysing every data set twice; once allowing \mcadam~to fit for a cluster imprint in the observed sky and once prohibiting the presence of any SZ signal. Furthermore, from the marginalised posterior probability distributions of each parameter we can extract constraints on the fitted parameters and hence conduct reliable parameter estimation which further enhances confidence in the detection. [Note that this $R$ value, suitable for use when a cluster $z$ is known, is \textit{different} from the $R$ value for blind detection discussed in \citet{Shimwell2012}.]

\section{Simulations}\label{sec:simulations}
In addition, to analysing the AMI SZ data, we also generate a cluster surface brightness profile using the X-ray fitted $\beta$-model estimates from \citet{Culverhouse2010} for each object. In Table \ref{tab:xray}, we list these parameters ($kT$, $L_X$, $n_{e,0}$, $r_c$, $\beta$) and also show the calculated $M_{\mathrm{T},200}$ and $f_{gas}$ using the mass-temperature relation from Equation (\ref{eq:MTrel}) and the ratio between $M_{\mathrm{gas},200}$ and $M_{\mathrm{T},200}$, respectively.

For each target, we insert the cluster profile on a modelled sky containing the point sources as measured by the LA, source confusion and primordial CMB. Finally, we create a simulated observation by using the AMI sampling function and instrumental noise, assumed Gaussian, to generate realistic \textit{uv}-data set. Table \ref{tab:sim-res} shows the recovered parameters and the integrated flux densities at 15.5 GHz for each simulated cluster. The latter are derived by simulating a typical observation with \textsc{profile}, excluding all astronomical background (point sources, CMB and confusion noise), to derive the net SZ flux taking into account the instrumental noise and sampling function of the SA.

It is evident that, when calculating the gas mass fraction from the at $r_{2500}$ derived $\beta$-profile parameters as well as the mass temperature scaling relation from equation \ref{eq:MTrel}, XMJ0849+4452 in particular has a very low predicted $f_{gas}$ compared to the other clusters (see Table \ref{tab:xray}), which is a symptom of three problems:
\begin{enumerate}

\item One issue arising from a small-radius fit to estimate the quantities internal to a larger radius is that any (small) error will cause large uncertainties on quantities calculated at $r_{200}$. Again, since the $n^2$-weighted X-ray temperature is more sensitive to clumping and shocking than the $n$-weighted SZ temperature measurement, such X-ray measurements are biased high.

\item Assuming isothermality with a temperature measured near the cluster core (excising any cooling flow), and taking the gas mass fraction to be constant throughout the cluster, will each introduce a bias in the mass estimates at high radius.
For example, \citet{Mroczkowski2011} found that the derivation of the total mass using SZ data and the virial theorem alone is probably most impacted by the assumption of constant gas mass fraction. 
\item Estimating the mass internal to a radius of $r_{2500}$ from X-ray data (using e.g. `onion peeling') can work well.
However, for instruments (such as AMI and SZA) that are sensitive to larger scales, a problem arises along the line-of-sight of $Y$ up to a fiducial radius~$r$, 
\begin{equation}
Y \equiv \int_\Omega y d\Omega = \frac{\sigma_{T}}{m_e c^2} \int_{-\infty}^{+\infty} dl \int_0^r n_e (r') T_e(r') 4 \pi r'^2 dr' \mathrm{.}
\end{equation} 
The measured SZ flux of the telescope is related to its sensitivity to a certain scale on the sky rather than to a defined radius, which means that the \textit{actual} line-of-sight contribution of $Y$ between the defined radius ($r_{2500}$) and $\pm \infty$, is not taken out when relating $Y$ to the cluster temperature and mass parameters because, for example, the true value of $\beta$ at large radius may be significantly different from the one assumed closer in. There is no easy way to correct for this. 
The SA has one advantage in these matters: because of its wider field of view and synthesised beam size, any measurement will give values of temperature and gas fraction that are averages internal to $r_{200}$ and hence deliver less biased (in these respects) estimates of the mass and temperature.

\end{enumerate}

Ideally, one would try to measure a radial $f_{gas}$ and $M_\mathrm{tot}$ profile via a joint analysis of different instruments sensitive to different radii. 
With this in mind, the following simulations and comparisons should be regarded as a rough estimation of cluster detectability rather than an accurate parameter comparison. Two targets in the sample (ISCS1438+34 and SPJ1638+4039) were not simulated as there were no X-ray fitted values nor any SZ mass measurements provided in \citet{Culverhouse2010}. For SPJ1638+4039, we use a mass estimate at $r_{200}$ derived from measured velocity dispersions (\citealt{Muzzin2009}) to give an educated guess on the detectability of this target. The fact that they are both clusters that have been successfully identified in IR but not using the SZ effect at radio frequencies hints at a complex radio environment or a cluster mass that is well below the thermal SZ detection limit of the observing telescope.

Our in-house package \textsc{profile} can simulate a patch of sky populated by a galaxy cluster, point sources and primordial CMB contribution. We use the
   source information from the LA raster observations to model the known
   sources in the field and add a population of fainter sources following the 10C source counts \citet{Davies2011}, which serve to simulate source confusion noise. \textsc{profile} lets us simulate the process of observing accurately by sampling in \textit{uv}-space according to the array's configuration, observing times and frequencies and adds instrumental noise (assumed Gaussian) to each visibility. The simulated \textit{uv}-data are then processed through the same data reduction and analysis pipeline as the real data. Since all of the simulations were generated using a $\beta$-model, it does not make sense to analyse them using another model (i.e. the DM-GNFW model) as a fitted $\beta$-model will always be the best match to our simulations and deliver the best evidence of detection. We also investigate the parameter constraints of each cluster to predict an estimated integrated SZ flux of the cluster. This analysis will contribute to estimating the feasibility of a real detection from the real data.

\begin{table*}
\begin{center}
\caption{Values from $\beta$-model X-ray fits taken from \citet{Culverhouse2010} that we use for the AMI simulations as well as $r_{200}$, the gas mass fraction, total mass at $r_{200}$ derived for these values using the mass-temperature scaling relation mentioned in Equation (\ref{eq:MTrel}).}

\label{tab:xray}
\begin{tabular}{lcccccccc|}
Cluster & k$T$ & L$_\mathrm{X}$ & $n_{e,0}$ & $r_{c}$ & $r_{200}$ & $\beta$ & $f_\mathrm{gas}$ & $M_\mathrm{T,200}$  \\
name & (keV) & ($\times 10^{44}$ erg s$^{-1}$) & $(10^{-2} cm^{-3})$ & (arcmin) & (arcmin) & (fixed) & & ($\times 10^{14} M_{\odot}$) \\
\hline \hline
CLJ1415+3612 & $6.5^{+0.9}_{-0.8}$ & $10^{+1}_{-1}$ & $2.25^{+0.14}_{-0.14}$ & $10.9^{+0.4}_{-0.4}$ & $1.39^{+0.1}_{-0.09}$ & 0.7 & $0.094^{+0.021}_{-0.019}$ & $5.63^{+1.17}_{-1.04}$  \\ [1mm]
XMJ0849+4452 & $6.7^{+2.0}_{-1.5}$ & $2.1^{+0.4}_{-0.4}$ & $0.67^{+0.08}_{-0.07}$ & $12.1^{+1.1}_{-1.0} $ & $2.06_{-0.23}^{+0.31}$ & 0.7 & $0.035^{+0.016}_{-0.013}$ & $5.17_{-1.76}^{+2.32}$
\\ [1mm]
RDJ0910+5422 & $4.5^{+1.5}_{-0.9}$ & $1.7^{+0.2}_{-0.2}$ & $0.65^{+0.09}_{-0.08}$ & $17.9^{+3.0}_{-1.7} $ & $1.87_{-0.19}^{+0.31}$ &0.7 & $0.101^{+0.064}_{-0.038}$ & $3.1_{-0.93}^{+1.55}$   \\ [1mm]
XMJ0830+5241 & 7.6$^{+0.8}_{-0.8}$ & 16$^{+1}_{-1}$ & 0.83$^{+0.03}_{-0.03}$ & 28.6$^{+1.0}_{-0.9}$ & $2.67_{-0.14}^{+0.14}$ & 0.7 & $0.181^{+0.037}_{-0.036}$ & $7.28^{+1.15}_{-1.15}$  \\ [1mm]
\end{tabular}
\end{center}
\end{table*}

\begin{table}
\begin{center}
\caption{Derived gas mass fraction and total mass at $r_{200}$ from the \mcadam~analysis for each simulated AMI observation. The X-ray derived parametres from \citet{Culverhouse2010} are used as input for the cluster simulation. We also present the estimated net SZ integrated flux density in the SA's 15.5-GHz channel calculated by the simulation software (see text).}\label{tab:sim-res}
\begin{tabular}{c|cc|c}
Cluster & $f_\mathrm{gas}$ & $M_\mathrm{T,200}$ & Calculated Integ. \\
name & & ($\times 10^{14} M_{\odot}$) &  flux density (\uJy) \\ \hline
CLJ1415+3612 & \multicolumn{2}{|c|}{not detected} & -715 \\
XMJ0849+4452 & \multicolumn{2}{|c|}{not detected} &-282 \\
RDJ0910+5422 & $0.124_{-0.03}^{+0.03}$ & $2.48_{-2.16}^{+0.52}$ & -426 \\
XMJ0830+5241 & $0.126_{-0.02}^{+0.02}$ & $9.77_{-1.56}^{+1.60}$ & -515 \\
\end{tabular}
\end{center}
\end{table}

\section{Results and Discussion}\label{sec:sa_results}

For each target, we discuss the source environment, the simulation of the observation conducted and hence the prospects of a successful cluster detection and source subtraction. In addition, we provide evidence ratios and present the 1D and 2D posterior marginals to formalise a detection. 

\begin{figure*}
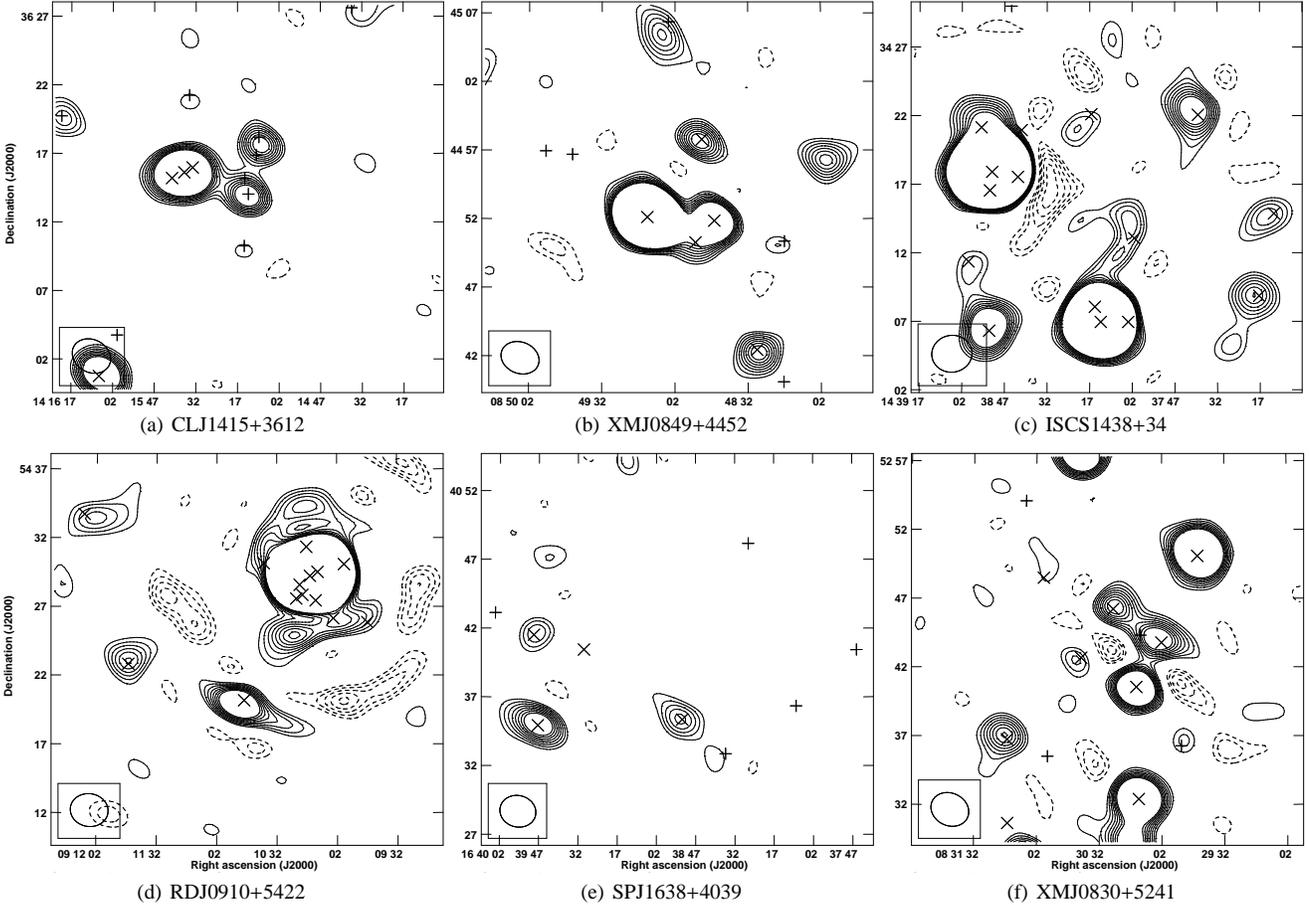


\subfigure[CLJ1415+3612]{\includegraphics[height=5.4cm, trim = 3mm 22mm 4mm 12mm, clip]{figures/CLJ1415_map}\label{subfig:CLJ1415_map}}
\subfigure[XMJ0849+4452]{\includegraphics[height=5.4cm, trim = 11mm 22mm 4mm 12mm, clip]{figures/XMJ0849_map}\label{subfig:XMJ0849_map}}
\subfigure[ISCS1438+34]{\includegraphics[height=5.4cm, trim = 11mm 22mm 4mm 12mm, clip]{figures/ISCS1438_map}\label{subfig:ISCS1438_map}}
\subfigure[RDJ0910+5422]{\includegraphics[height=5.6cm, trim = 3mm 16.2mm 4mm 12mm, clip]{figures/RDJ0910_map}\label{subfig:RDJ0910_map}}
\subfigure[SPJ1638+4039]{\includegraphics[height=5.6cm, trim = 11mm 16.2mm 4mm 12mm, clip]{figures/SPJ1638_map}\label{subfig:SPJ1638_map}}
\subfigure[XMJ0830+5241]{\includegraphics[height=5.6cm, trim = 11mm 16.2mm 4mm 12mm, clip]{figures/XMJ0830_map}\label{subfig:XMJ0830_map}}

\caption{SA maps of the cluster sample. The contours are scaled linearly in integer multiples of the map noise level found in table \ref{tab:observations}, starting at the 3-$\sigma$ level. Negative levels are shown with dashed lines and positive ones have solid contours. The synthesised beam is depicted on the lower LHS. The positions of the sources found by the LA (see section \ref{sec:sourcefind}) are displayed with `$\times$'s if the individual integrated source flux density is higher than 4 times the SA noise level and a `+' for fainter sources (see section \ref{sec:mcadam}).}\label{fig:maps}

\end{figure*}

\begin{figure*}
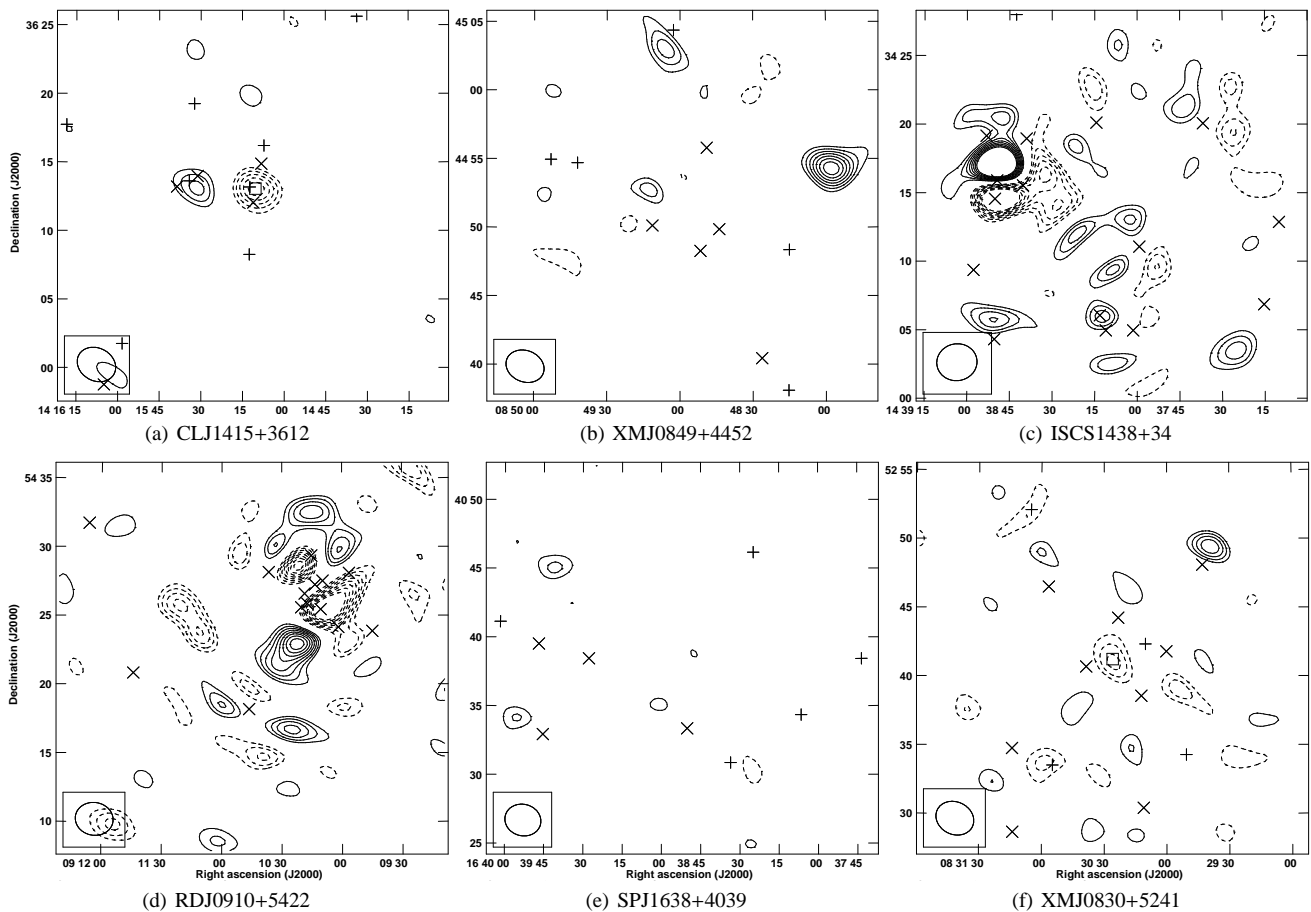

\subfigure[CLJ1415+3612]{\includegraphics[height=5.4cm, trim = 3mm 22mm 4mm 12mm, clip]{figures/CLJ1415s_map}\label{subfig:CLJ1415s_map}}
\subfigure[XMJ0849+4452]{\includegraphics[height=5.4cm, trim = 11mm 22mm 4mm 12mm, clip]{figures/XMJ0849s_map}\label{subfig:XMJ0849s_map}}
\subfigure[ISCS1438+34]{\includegraphics[height=5.4cm, trim = 11mm 22mm 4mm 12mm, clip]{figures/ISCS1438s_map}\label{subfig:ISCS1438s_map}}
\subfigure[RDJ0910+5422]{\includegraphics[height=5.6cm, trim = 3mm 16.2mm 4mm 12mm, clip]{figures/RDJ0910s_map}\label{subfig:RDJ0910s_map}}
\subfigure[SPJ1638+4039]{\includegraphics[height=5.6cm, trim = 11mm 16.2mm 4mm 12mm, clip]{figures/SPJ1638s_map}\label{subfig:SPJ1638s_map}}
\subfigure[XMJ0830+5241]{\includegraphics[height=5.6cm, trim = 11mm 16.2mm 4mm 12mm, clip]{figures/XMJ0830s_map}\label{subfig:XMJ0830s_map}}
\caption{Source-subtracted maps of the cluster sample observed by the SA. The contours are scaled linearly in integer multiples of the map noise level found in table \ref{tab:observations}, starting at the 3-$\sigma$ level. Negative levels are shown with dashed lines and positive ones have solid contours. The synthesised beam is depicted on the lower LHS. The positions of the subtracted sources are displayed with `$\times$'s if their individual integrated flux densities are higher than 4 times the SA noise level and a `+' for fainter sources (see section \ref{sec:mcadam})}.\label{fig:maps_s}

\end{figure*}

We show the maps of the SA observations in Fig. \ref{fig:maps} and the source-subtracted maps after Bayesian Analysis in Fig. \ref{fig:maps_s} for each target. CLJ1415+3612 is presented first because it is the most confident detection, followed by XMJ0849+4452, a prime example of non-detection due to source contamination and how the analysis software behaves in such a case.

\subsection*{CLJ1415+3612}

This cluster has a $\approx$1.3 mJy source located directly on top of it, a 0.9~mJy source $\approx 1.2 \arcm$ to the north and a 0.5/0.6 mJy pair $\approx 3 \arcm$ to the north. Moreover, three other sources lie about 5\arcm~to the east with integrated flux densities 0.7, 0.8, 0.2 mJy respectively; the central source of the three could be matched in both SDSS and 2MASS data to NGC 5529, a galaxy in a group at redshift 0.009 (\citealt{Kochanek2001}). Fig. \ref{subfig:CLJ1415_map} shows the SA map. We note as well that the integrated fluxes of the three sources summed together are lower on the LA map than the unresolved single feature on the SA map in the same location. This is evidence for extended emission, which would be plausible for a close-by galaxy or group of galaxies. As these sources are comparatively low in flux, we managed to successfully subtract the sources from the data using our fitted values and recover a $7$-$\sigma$ decrement with an integrated flux of $\approx  780$~\uJy (Fig. \ref{subfig:CLJ1415s_map}), which compares to a $6$-$\sigma$ decrement on the SZA. Given the lower noise level of the AMI observation but also the fact that the SZ effect is approximately 3.5 times stronger at 30 GHz than at 15 GHz, our expected signal-to-noise ratio for the SA with present noise level inferred from the SZA map would only be $\approx 2.5\sigma$. The difference arises primarily from the array configurations; the SZA has eight, 3.5-m dishes in a closed-packed configuration with baselines of $350\lambda-1300\lambda$ providing a resolution of $\approx 2\arcmin$ at 30 GHz compared to the 10-dish AMI SA ($200\lambda - 1000\lambda$ baselines, $3\arcm$ resolution). For a more detailed comparison between the SZA and AMI, see \citet{Shimwell2012b}.

Judging from the simulations, we do not expect to detect this cluster. Although the net integrated flux density on AMI channel 5 from the thermal SZ effect created by the simulated cluster is about $715$ \uJy (which is in agreement with the observed data), there is a point source (with flux density $\approx$1.3 mJy) directly at the phase centre and a few more radio sources within 3\arcm of the cluster position. As the simulated cluster has an $r_{200} \approx 2.39$\arcm, it will be point source-like in the simulated SA observation and therefore challenging for \mcadam~to disentangle from data contaminants. We consider this to be the main reason for the non-detection; the SZ imprint generated from the at $r_{2500}$ fitted $\beta$-profile is too small and hence the simulated cluster will have a very peaked and steep profile. This makes it very difficult for the SA to recover it. Thus, the evidence ratio in favour of a non-detection and the parameters are not constrained by the simulated data, which can be further attributed to the simulated cluster's small gas mass fraction and core radius, which leads to a small derived gas mass ($\approx 5.3 \times 10^{13} M_\odot$) and hence to a faint thermal SZ effect.
In complete contrast, the AMI data show a clear SZ detection: the model selection ratios for each parameterisation indicate a decisive detection ($R > 8$) which is substantiated further by the good mass constraints of each model, presented in Table \ref{tab:m200}. Both measured masses are higher than the calculated one, the $\beta$-model estimates a mass of $\approx 7.3 \times 10^{14} M_\odot$ which agrees with $\approx 10.4 \times 10^{14} M_\odot$ from the DM-GNFW model within $2\sigma$. This is evidence for the low-radius ($r_{2500}$) derived SZ profile being inconsistent with the actual observed one, which is measured at $r_{200}$.

\begin{figure}
\includegraphics[width=7cm]{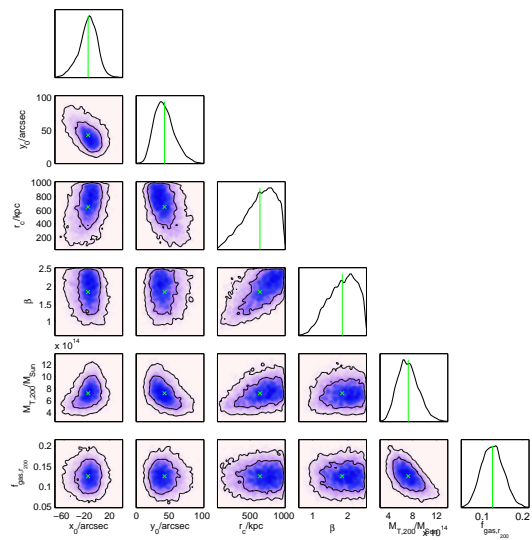}

\caption{1D and 2D marginalised posterior distributions for CLJ1415+3612 ($\beta$-model). $x_0$/$y_0$ are the offsets in Arcseconds in R.A./DEC. from the observation's phase centre. The green crosses and lines show the mean of the respective marginal distribution. The contours on the 2D marginals indicate the areas enclosing 68\% and 95\% of the probability distribution.}\label{fig:CLJ1415_tri}
\end{figure}

\begin{figure}
\includegraphics[width=7cm]{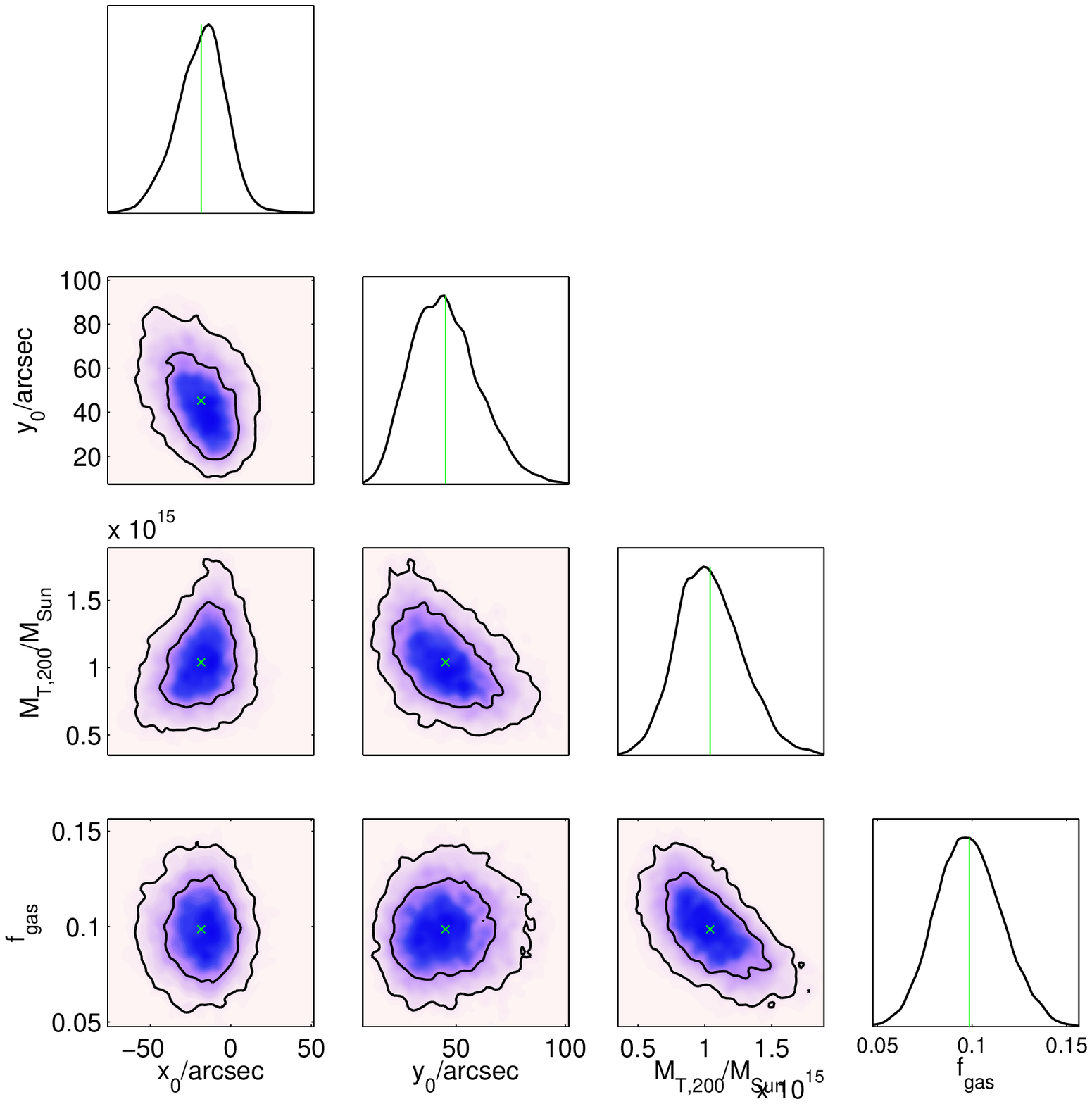}

\caption{1D and 2D marginalised posterior distributions for CLJ1415+3612 (DM-GNFW model). $x_0$/$y_0$ are the offsets in Arcseconds in R.A./DEC. from the observation's phase centre. The green crosses and lines show the mean of the respective marginal distribution. The contours on the 2D marginals indicate the areas enclosing 68\% and 95\% of the probability distribution.}\label{fig:CLJ1415a_tri}
\end{figure}

We show the constraints on each parameter of the cluster model via their 1D and 2D marginalised posterior distributions for the $\beta$-model (Fig. \ref{fig:CLJ1415_tri}) and (Fig. \ref{fig:CLJ1415a_tri}) for the DM-GNFW model. It is also well known that the $\beta$-model exhibits strong degeneracies between the parameters $\beta$ and $r_c$, which is reflected by the fact that the 2D-marginalised posterior for these two parameters stays ill-constrainted along the diagonal of the $r_c$-$\beta$ plane and also hits the upper prior edges.

\subsection*{XMJ0849+4452}

This field suffers from substantial contamination from bright sources. The brightest ($\approx 10$ mJy) is located 2.75\arcmin~to the south-east, followed by an  approximately $1.6$ \mJy~source 3.3\arcm~to the south-west and a~$\approx 1$~mJy source to the north-west about 4\arcm~away.
The simulation of this cluster shows how the central bright source obstructs the cluster decrement and the analysis of the simulation does not predict a convincing detection (R=0.05).

\begin{figure}
\includegraphics[width=8cm, trim = 4mm 17mm 4mm 12mm, clip]{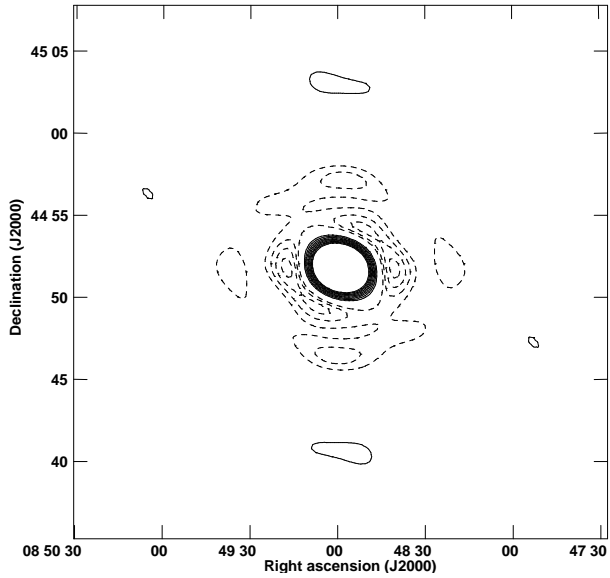}
\caption{Synthesised beam of the SA for XMJ0849+4452. Contours start at the 6\% level for both positive (solid lines) and negative (dashed lines) scales and increase/decrease in 3 per cent steps.}\label{fig:XMJ0849_SB}
\end{figure}

\begin{figure}
\includegraphics[width=7cm]{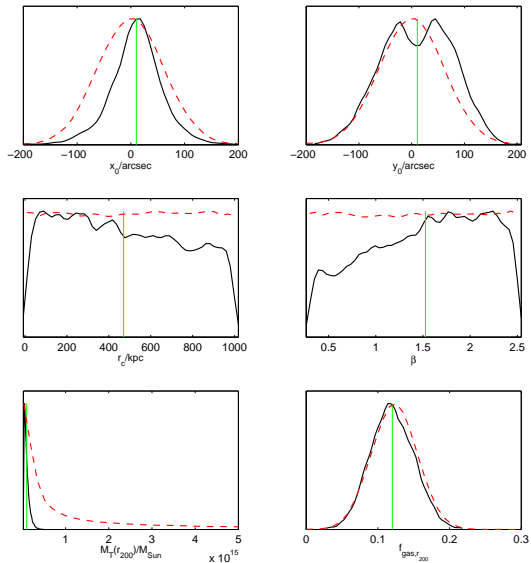}
\caption{1D marginalised posterior distributions (black) for the parameters of XMJ0849+4452 analysed with the $\beta$--model. $x_0$/$y_0$ are the offsets in Arcseconds in R.A./DEC. from the observation's phase centre. The red dashed lines display the prior distribution for each parameter and hence show the influence on each posterior. The green bars show the mean of each posterior distribution.}\label{fig:XMJ0849_1D}
\end{figure}

From the AMI data, our evidence ratios $R$ are  $< 0$ for both models. We present the 1D posterior marginals in Fig. \ref{fig:XMJ0849_1D} for the $\beta$--model to provide an example of what typical marginalised posterior distributions look like for a non-detection. Since the data do not contain any information to constrain the cluster parameters, we only recover the prior distributions or are heavily biased by them. To illustrate this point, we added the prior distributions in red in Fig.\ref{fig:XMJ0849_1D}; all of total mass, $r_c$ and $\beta$ are fully prior-driven in this example.

\subsection*{ISCS1438+34}
This observation is strongly contaminated by two bright sources (for a map, see Fig. \ref{subfig:ISCS1438_map}), one with an integrated flux of about 38~mJy, approximately $8.5$\arcm \,to the east (14 38 49,+34 16 00) and the another one, with integrated flux density of about 18 mJy, $\approx 9$\arcm~to the south (14 38 11, +34 05 06). Both sources are well embedded within a crowded environment and produce significant sidelobes. Although none of them are classified as extended on the LA, the parameter space to be explored in order to model them accurately is too large for \mcadam. If we fix all the source fluxes and spectral indices using delta priors from the LA measurements, apart from a few sources closest to the phase centre and deemed to cause the highest nuisance to the data, we reduce the dimensionality of the Bayesian fit, which improves convergence on a best fit in the analysis. However, we are sill not able to recover an SZ decrement nor any parameter estimates from the data. The model selection criterion strongly favours a non-detection. 

\subsection*{RDJ0910+5422}
Two sources with integrated fluxes of 36 and 17 mJy respectively on the LA map produce important sidelobes which swamp the entire map (Fig. \ref{subfig:RDJ0910_map}). The two sources form one extended feature on the SA map 7\arcm~to the north-west of the phase centre.
 
\begin{figure}
\begin{centering}
\includegraphics[width=7cm, trim = 6mm 17mm 4mm 12mm, clip]{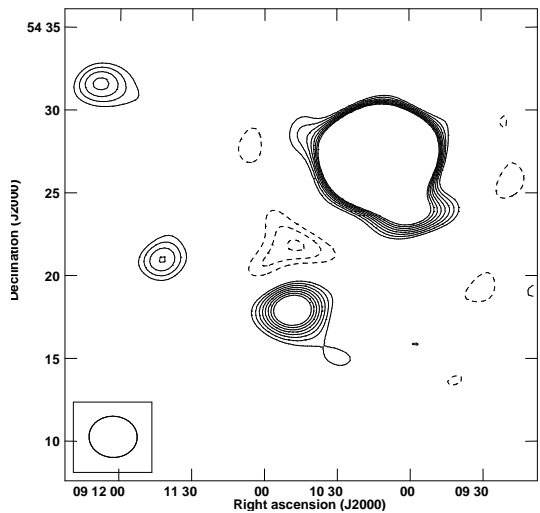}
\caption{Map of the simulated obsevation of RDJ0910+5422. The contours are scaled linearly in multiples of the noise level, starting at the 3$\sigma$-level. Dashed lines depict negative flux levels and solid contours show positive levels.}\label{fig:RDJ0910sim_map}
\end{centering}
\end{figure}

We do not expect this cluster to be detectable; Table \ref{tab:xray} indicates a relatively low-mass cluster with a low predicted integrated flux estimate (in Table \ref{tab:sim-res}). In the simulation however, we do manage to successfully recover some SZ flux despite a moderately successful source fit and subtraction. Judging from the map of the simulated data, shown in Figure \ref{fig:RDJ0910sim_map},  We expect a negative peak flux from the cluster of $\approx -400$ \uJy in the real data, which is less than 1 per cent of the brightest feature in the map and will be a potential issue for the dynamic range of the telescope due to correlator errors and residuals left from the source subtraction. This makes a successful SZ signal recovery and source-subtraction from real data very doubtful and indeed the negative evidence ratios ($-3.2$ for the $\beta$-model and $-2.9$ for the DM-GNFW model) confirm non-detections for this cluster.

\subsection*{SPJ1638+4039}
This target has a relatively clean source environment (Fig. \ref{subfig:SPJ1638_map}); the closest source from the phase centre is $\approx 5$\arcm~away and has an integrated flux of about 1 mJy. If we use the estimated mass at $r_{200}$ derived from measured velocity dispersions of SPJ1638+4039 (\citealt{Muzzin2009}), $M_\mathrm{200}~=~(2.4 \pm 1.8) \times 10^{14} M_\odot$ and the fact that this cluster is at high redshift ($z = 1.2$), we do not believe that this cluster is able to produce an SZ signal which is above our detection threshold. A simulation of the cluster's imprint on an empty sky, using the DM-GNFW model and assuming a gas mass fraction of 0.123 gives an integrated flux estimation of $\approx  240$~\uJy, which would be a 2-$\sigma$ detection.
In the SA data, the source subtracted map (Fig. \ref{subfig:SPJ1638s_map}) does not show any evidence of SZ flux and reflected by a negative model selection parameter for both models in Table \ref{tab:evidences}.

\subsection*{XMJ0830+5241}
Fig. \ref{subfig:XMJ0830_map} shows the SA map of XMJ0830+5241, which has a moderately crowded source environment. However all the closest sources, within a distance of 1 synthesised beam, are relatively low in flux (1 mJy and less). There is another source about 3.5\arcm~to the south-west, which has an integrated flux density of $\approx 2.2$ mJy. Furthermore significant decrements can be seen on the map, prior to any source subtraction which cannot be assigned to the sidelobes of any radio sources in the field. Using the reference synthesised beam from Fig. \ref{fig:XMJ0849_SB} we see that only a maximum of $\approx 120$ \uJy can be attributed to the $\approx 2.2$ mJy source, since the first sidelobe to the north-east of the synthesised beam is about 6\%. However the decrement has an integrated flux density of about 360 mJy. Each observation of XMJ0830+5241 on the SA has a secondary calibrator interleaved every 6 minutes. When phase-calibrating the data in \textsc{reduce}, we carry out phase stability tests on each channel for every baseline and reject data with poor calibration, which we assess by estimating the phase errors for each 400s visit to the astrometrical calibrator and rejecting it if the error is greater than 15\deg~(the calculated typical error is about 4$\deg$~per channel) or if the phase step of two consecutive calibrator observations is more than 30$\deg$, thus ensuring that phase errors in the data are small and could have not conspired to create this $5$-$\sigma$ SZ effect.

\begin{figure}
\includegraphics[width=7cm]{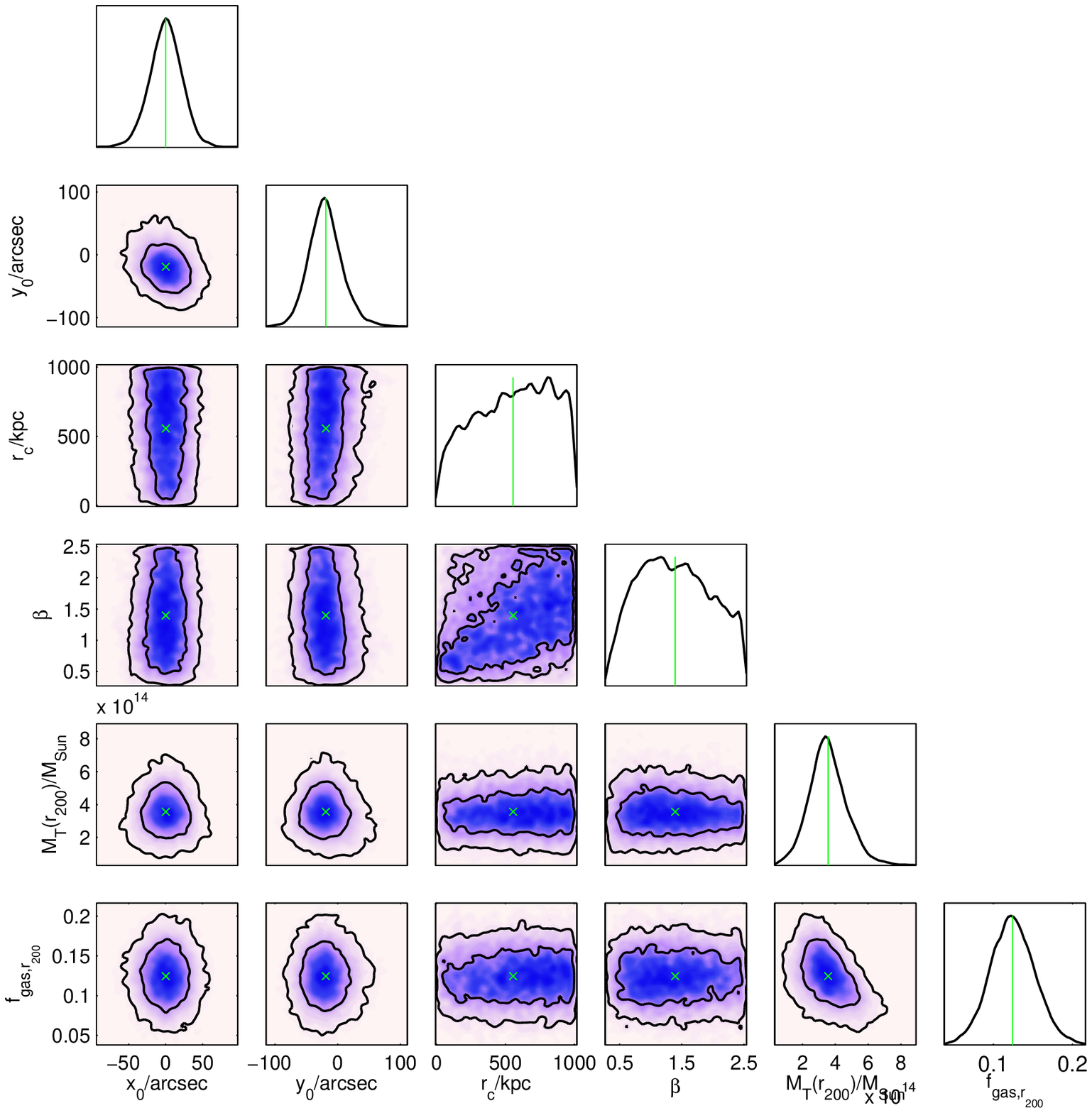}
\caption{1D and 2D marginalised posterior distributions for XMJ0830+5241 ($\beta$-model). $x_0$/$y_0$ are the offsets in Arcseconds in R.A./DEC. from the observation's phase centre. The green crosses and lines show the mean of the respective marginal distribution.The contours on the 2D marginals indicate the areas enclosing 68\% and 95\% of the probability distribution.}\label{fig:XMJ0830_tri}
\end{figure}

\begin{figure}
\includegraphics[width=7cm]{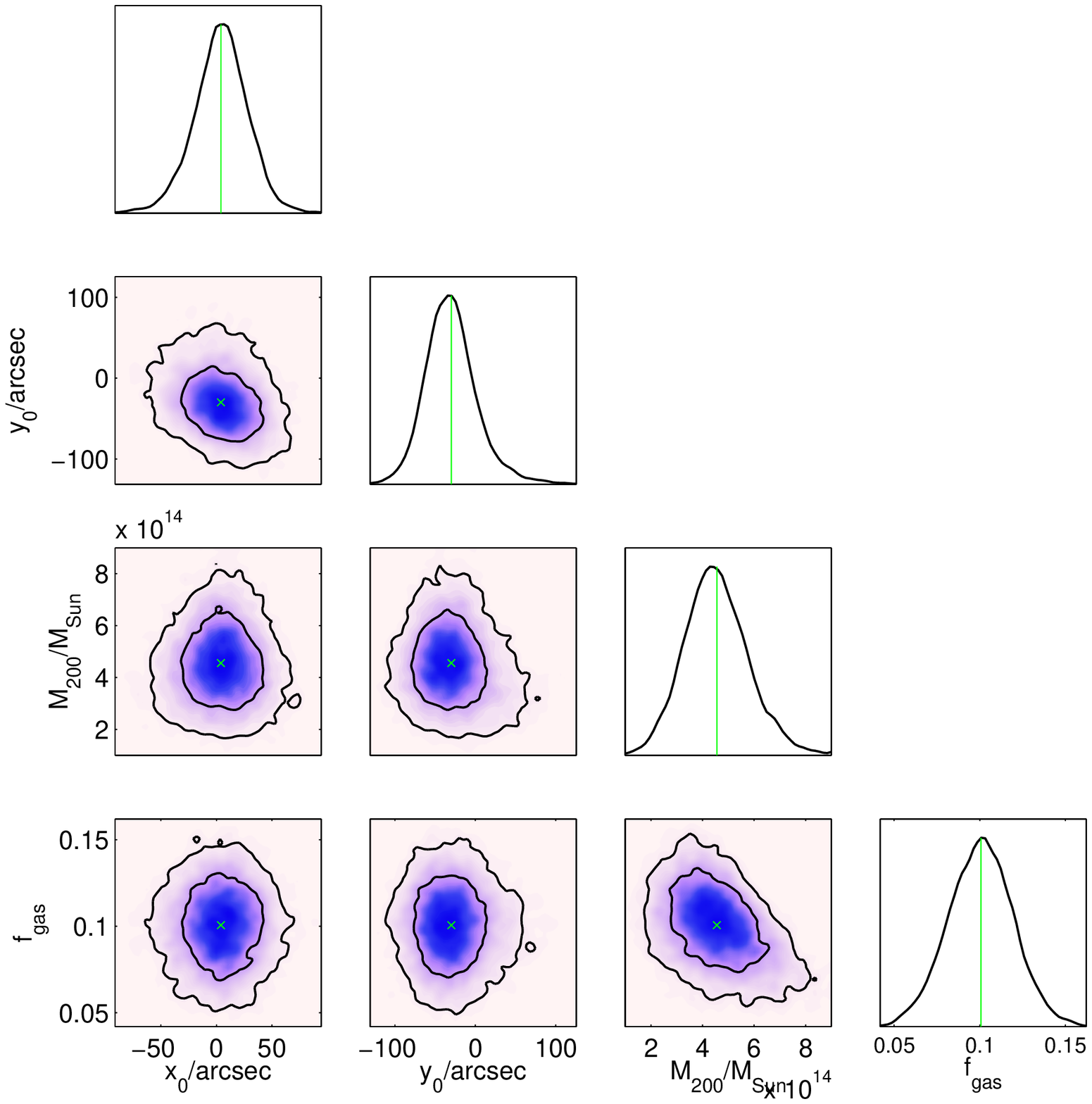}

\caption{1D and 2D marginalised posterior distributions for XMJ0830+5241 (DM-GNFW model). $x_0$/$y_0$ are the offsets in Arcseconds in R.A./DEC. from the observation's phase centre. The green crosses and lines show the mean of the respective marginal distribution. The contours on the 2D marginals indicate the areas enclosing 68\% and 95\% of the probability distribution.}\label{fig:XMJ0830a_tri}
\end{figure}

Simulating this observation using the source fluxes and spectral indices derived from LA data (section \ref{sec:sourcefind}) as well as the cluster parameters from Table \ref{tab:xray} strongly favours a detection ($R = 3$), which is also expected when mapping the simulated \textit{uv}-data in which the cluster's imprint appears with more than 6 times the noise level of the observation (For channel 5, the predicted flux density (shown in Table \ref{tab:sim-res}) is $-515$\uJy. This signal-to-noise ratio agrees with the $6$-$\sigma$ decrement found by \citet{Culverhouse2010}. However, when extrapolating from the SZA measurement taking into account the difference in noise level of each observation as well as the spectral intensity difference of the thermal SZ effect between 30 and 15 GHz, but ignoring the difference in array configuration, we would expect the decrement to only appear at a $\approx 4$-$\sigma$ level in the AMI data rather than the measured $6$-$\sigma$.
Analysing the SA data, the evidence ratio for a detection is substantial for both parameterisations ($R >$3) although less strong than in the case of CLJ1415+3612. After source subtraction, the integrated flux of the the cluster's imprint is $\approx 600$ \uJy. The posterior marginals, shown in Fig.~\ref{fig:XMJ0830_tri} for the $\beta$-model and in Fig.~\ref{fig:XMJ0830a_tri}) for the DM-GNFW model, show good constraints on the mass for both parameterisations; they agree with each other within one $\sigma$ and indicate a relatively low-mass cluster of $\approx~4~\times~10^{14}~M_\odot$. Again, the degeneracies between $\beta$ and $r_c$ for the isothermal $\beta$-model limit a tight constraint on the joint 2D posterior marginal of these parameters and result in the distribution hitting the upper prior edges.



\begin{table}

\caption{Evidence ratios R for each cluster and different parameterisations. Also provided are the evidence ratios for the simulations.}\label{tab:evidences}
\begin{tabular}{|ccc|c|}

Cluster &  $\beta$-model & DM-GNFW model & simulation  \tabularnewline
\hline
\hline
CLJ1415+3612 &  8.6 & 8.0 & -0.1 \tabularnewline
XMJ0849+4452 & -0.64 & -0.2 & 0.05 \tabularnewline
ISCS1438+34  & -6.1 & -5.4 & no data \tabularnewline
RDJ0910+5422 & -3.2 & -2.9 & 0.6 \tabularnewline
SPJ1638+4039 & -1.1 & -1.9 & no data \tabularnewline
XMJ0830+5241 & 3.0 & 3.2 & 4.3 \tabularnewline

\hline
\end{tabular}
\end{table}

\begin{table}
\caption{Mean values and their 68\% confidence limits of $M_\mathrm{T, 200}$ for each cluster and parameterisation.} \label{tab:m200}
\begin{tabular}{ccc}
Cluster & $M_\mathrm{T, 200} / M_\odot$ & $M_\mathrm{T, 200} / M_\odot$ \\
name & ($\beta$-model, SA data) & (DM-GNFW, SA data) \\
\hline \hline
CLJ1415+3612 & $7.27^{+1.78}_{-1.77} \times 10^{14}$ & $10.40_{-2.36}^{2.45} \times 10^{14}$ \\[1mm]
XMJ0830+5241 & $3.56^{+1.10}_{-1.11} \times 10^{14}$ & $4.66_{-1.41}^{+1.44} \times 10^{14}$ \\
\end{tabular}
\end{table}

\section{LA SZ measurements}\label{sec:la_followup}
For our two SA SZ detections, XMJ0830+5241 and CLJ1415+3612, we carry out additional follow-up observations with the LA. As both clusters are at high redshift (0.99 and 1.03 respectively) we aim to investigate whether these clusters' angular extent would be better geared to the baseline range of the LA and hence improve parameter estimations. Note that our Bayesian analysis expects the cluster's imprint to be an extended feature and hence have a different profile as a function of \textit{uv}-distance than a point source.  
From our SA analysis, we find $r_{200}\approx 2.3$\arcm and $r_{200} \approx 2.9$~\arcm for CLJ1415+3612 and XMJ0830+5241, respectively. As the angular resolution of the LA is $\approx 30 \arcs$ using all of its baselines, these two clusters will be resolved by the LA. However, the trade-off is that the LA cannot access the \textit{uv}-points that have the most significant SZ signal.

\begin{figure*}
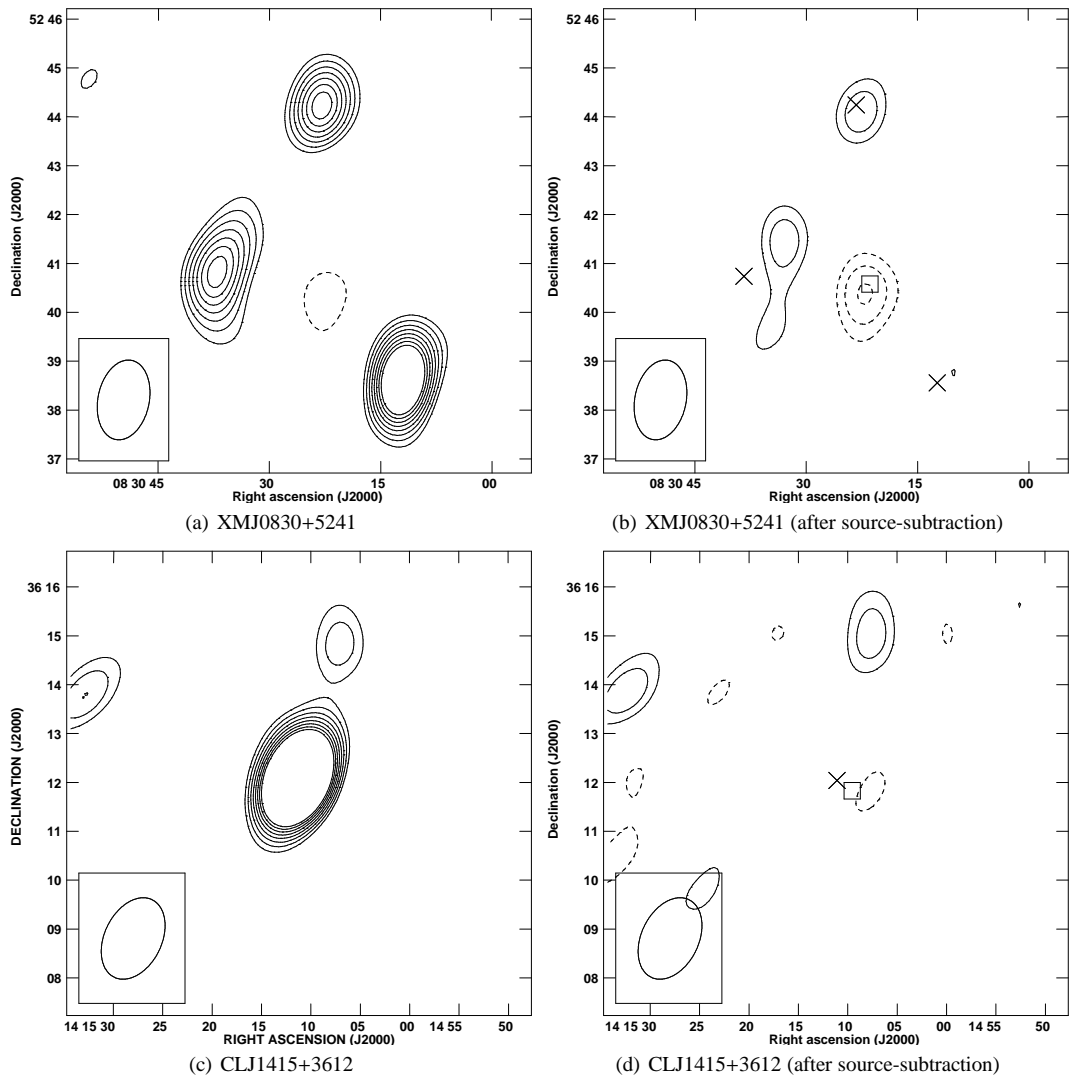

\subfigure[XMJ0830+5241]{\includegraphics[width=7cm, trim = 3mm 17mm 4mm 12mm, clip]{figures/XMJ0830_LAmap}
\label{subfig:XMJ0830_LAmap}}
\subfigure[XMJ0830+5241 (after source-subtraction)]{\includegraphics[width=7cm, trim = 3mm 17mm 4mm 12mm, clip]{figures/XMJ0830s_LAmap}
\label{subfig:XMJ0830s_LAmap}}
\subfigure[CLJ1415+3612]{\includegraphics[width=7cm, trim = 3mm 17mm 4mm 12mm, clip]{figures/CLJ1415_LAmap}
\label{subfig:CLJ1415_LAmap}}
\subfigure[CLJ1415+3612 (after source-subtraction)]{\includegraphics[width=7cm, trim = 3mm 17mm 4mm 12mm, clip]{figures/CLJ1415s_LAmap}
\label{subfig:CLJ1415s_LAmap}}
\caption{Maps of the deep single pointed observations of XMJ0830+5241 (top) and CLJ1415+3612 (bottom) on the LA before (left) and after (right) source subtraction. The mapping was done using a \textit{uv}-taper of 1.2 k$\lambda$. Crosses show the positions of subtracted sources and the box indictates the position of the cluster found by \mcadam~in the SA data. Contours scale in integer multiples of the noise level, starting at $3$-$\sigma$; negative contours are displayed with dashed lines while solid ones denote positive ones. The synthesised beam is displayed as an inlaid on the lower left-hand side of the map.}\label{fig:LAmaps}
\end{figure*}

We note that the maps agree with what we expect; the LA map of XMJ0830+5241 (Fig. \ref{subfig:XMJ0830_LAmap}) show a $3$-$\sigma$ decrement which agrees with the SA position. After source-subtraction (Fig. \ref{subfig:XMJ0830s_LAmap}), we manage to recover a $5\sigma$ decrement.
The deep follow-up CLJ1415+3612 is swamped by the $\approx 1.3$mJy located on top of the cluster position, as shown in Fig. \ref{subfig:CLJ1415_LAmap}. After source-subtraction, there is a small, just over $3$-$\sigma$, decrement in the source-subtracted map whose position agrees with the SA observation.

We cannot do any Bayesian analysis of the LA data with our current models, which require application of the virial theorem internal to $r_{200}$, because the LA is not able to measure up to an angular scale which corresponds to $r_{200}$. The current parameterisation in \mcadam, however, uses $M_{\mathrm{T}, 200}$ as sampling parameter, hence, unconstrained $M_{\mathrm{T}, 200}$ or $r_{200}$ measurements result in any other profile parameters remaining unconstrained. Although \mcadam\,recovers positions in both cases which seem to agree with the locations of the respective cluster, the results of this analysis remain questionable as no detection probabilities nor parameter estimates could be derived.

\section{Conclusions}
AMI has observed a declination-limited subset of six of the \citet{Culverhouse2010} sample. 
\begin{enumerate}
\item Three of these targets (ISCS1438+34, RDJ0910+5422, XMJ0849+4452) suffered from heavy radio source contamination which inhibited a detection. Using a mass estimate from \citep{Muzzin2009} of SPJ1638+4039 and simulating a typical AMI-SA observation of the target, which predicts a thermal SZ effect that is below the detection threshold of our telescope. Hence, we believe the mass of SPJ1638+4039 to be too small to be detected by the SA.
\item Two are firm detections (CLJ1415+3612, XMJ0830+5241) with the SA, with Bayesian evidence ratio $R$ of $\approx 8$ and $\approx 3$ respectively. In the case of CLJ1415+3612, AMI is able to find a $7\,\sigma$ detection compared to $6\,\sigma$ on the SZA despite the thermal SZ effect being a factor of $\approx 3.5$ fainter at AMI's frequency band. Similarly, the integrated flux density of XMJ0830+5241 is higher than expected in the SA data ($\approx 360$\uJy) compared to the SZA measured integrated flux density, if only taking into account the difference in intensity of the thermal SZ effect between 30 and 15 GHz.

\item In the subset of six, AMI detects in SZ the same clusters (CLJ1415+3612, XMJ0830+5241) as \citet{Culverhouse2010} with the SZA. The AMI data and analysis, however, return parameters internal to $r_{200}$ ($\approx$ virial radius). For CLJ1415+3612 and XMJ0830+5241 respectively, our $\beta$-model, route finds $M_{\mathrm{T},200} = 7.3 \pm 1.8 \times 10^{14} M_\odot$ and $M_{\mathrm{T},200} = 3.6 \pm +1.1 \times 10^{14} M_\odot$, while our DM-GNFW model finds $M_{\mathrm{T},200} = 4.7 \pm 1.4 \times 10^{14} M_{\odot}$ and $M_{\mathrm{T},200} = 10.4 \pm 2.5 \times 10^{14} M_\odot$.

\item In using the X-ray data internal to $r_{2500}$ from \citet{Culverhouse2010} to simulate what AMI should see at larger radius, we have highlighted three causes of bias with respect to reality, when deriving parameters from SZ measurements if the modelling relies on X-ray data from the central region of the clusters.

\item We find SZ effects in the higher-resolution follow-up observation of CLJ1415+3612 and XMJ0830+5241 carried out with AMI LA.
\end{enumerate}
\section*{Acknowledgments}
We thank the anonymous referee for providing us with very constructive
comments and suggestions. We thank the staff of the Mullard Radio Astronomy Observatory for their invaluable assistance in the commissioning and operation of AMI, which is supported by Cambridge University and the STFC. MPS and CR gratefully acknowledge the support of STFC studentships and YCP gratefully acknowledges the support of a CCT/Cavendish Laboratory studentship. The analysis work was conducted on the Darwin Supercomputer of the University
of Cambridge High Performance Computing Service supported by HEFCE and the Altix 3700
supercomputer at DAMTP, University of Cambridge supported by HEFCE and STFC. We are grateful to Stuart
Rankin and Andrey Kaliazin for their computing assistance. This research has made use of the SIMBAD database, operated at CDS, Strasbourg.

\bsp

\label{lastpage}
\end{document}